\documentclass[aps,prstab,reprint,superscriptaddress]{revtex4-1}
\usepackage{graphicx}
\begin{document}

%\preprint{}

\title{Helical Undulator Based on\\Partial Redistribution of Uniform Magnetic Field}%


\author{N.~Balal}%
%\thanks{}
\affiliation{Ariel University, Ariel, Israel}%

\author{I.~V.~Bandurkin}%
 \email{iluy@appl.sci-nnov.ru}%
\affiliation{Institute of Applied Physics, Russian Academy of Sciences, Nizhny Novgorod, Russia}%

\author{V.~L.~Bratman}%
\affiliation{Ariel University, Ariel, Israel}
\affiliation{Institute of Applied Physics, Russian Academy of Sciences, Nizhny Novgorod, Russia}%

\author{A.~E.~Fedotov}%
\affiliation{Institute of Applied Physics, Russian Academy of Sciences, Nizhny Novgorod, Russia}%


\date{\today}

\begin{abstract}
A new type of helical undulator based on redistribution of magnetic field of a
    solenoid by ferromagnetic helix has been proposed and studied both in
    theory and experiment. Such undulators are very simple and efficient for
    promising sources of coherent spontaneous THz undulator radiation from dense
    electron bunches formed in laser-driven photo-injectors.
\end{abstract}

%\keywords{}

\maketitle

\section{Introduction}

Advanced laser-driven photo-injectors make possible formation of very dense
picosecond and sub-picosecond electron bunches with charge of the order of 1~nC
and moderate energy of 3--6~MeV \cite{Power_Inproceedings_2010_Overview,
Bartnik_PhysRevST-A_2015_Operational, Stephan_PhysRevST-A_2010_Detailed,
Rosenzweig_NuclInstMethPhysResA_2011_Design, Perez_ApplPhysLett_2014_High}.
Such bunches can be attractive for production of power THz electromagnetic
pulses \cite{Doria_IEEEJQE_1993_Coherent, Gover_PhysRevLett_1994_Time,
Lurie_PhysRevST-A_2007_Enhanced, Lurie_PhysRevST-A_2015_Single,
Bratman_NuclInstMethPhysResA_2001_Generation,
Lee_NuclInstMethPhysResA_2015_Numerical, Balal_ApplPhysLett_2015_Negative,
Lurie_PhysRevST-A_2016_Energy}, if very strong mutual Coulomb longitudinal
repulsion of bunch particles is prevented or mitigated.  One of the recently
proposed efficient methods for stabilization of the electron bunches length and
for corresponding increase of their spontaneous Doppler-up-shifted terahertz
undulator radiation energy \cite{Balal_ApplPhysLett_2015_Negative,
Lurie_PhysRevST-A_2016_Energy} is based on development of the so-called
Negative-Mass Instability (NMI) \cite{Freund_PhysRevA_1983_Unstable,
Ganguly_PhysRevA_1985_Nonlinear, Ginzburg_TechPhys_1988_Non,
Freund_ISBN0_1995_Principles} in the bunches moving in undulators.
Implementation of the method requires the undulator field to be combined with a
strong uniform axial guiding magnetic field
\cite{Balal_ApplPhysLett_2015_Negative, Lurie_PhysRevST-A_2016_Energy}.  Here,
the ``strong'' means that the value of the uniform guiding magnetic field
should be higher than the resonant value corresponding to the coincidence of
undulator and cyclotron frequencies of the electrons. In this paper, we
demonstrate that in such a system, the undulator field can be easily created by
redistributing the strong uniform field by a periodic ferromagnetic helix.
A similar method was proposed and demonstrated many years ago in planar undulator
systems with periodic planar insertions
\cite{Ho_NuclInstMethPhysResA_1990_Novel,
Varfolomeev_NuclInstMethPhysResA_1993_Strong, Bratman_Inproceedings_1998_With}.
Also, helically distributed planar elements were used for creating a relatively
low helical transverse magnetic field
\cite{Ohigashi_NuclInstMethPhysResA_2003_Construction}.  We suggest using a
more convenient helical undulator based on a fully-symmetrical helical
insertion \cite{Balal_Inproceedings_2015_Efficient}.  It is fairly obvious that
magnetization of a ferromagnetic helix in the guiding field provides a helical
component of the magnetic field, and we will show that a sufficient for
realization of the NMI-based terahertz source amplitude of this component can
be readily obtained.

\section{Analytical estimating}

According to \cite{Balal_ApplPhysLett_2015_Negative,
Lurie_PhysRevST-A_2016_Energy}, development of NMI inside a dense electron bunch can
lead to effective mutual attraction of particles. As a result, a dense core
with a fairly stable longitudinal size smaller than the radiation wavelength
can appear. Such longitudinal bunching occurs in a combined helical,
$\vec{B}_u$, and over-resonance uniform guiding, $B_0\vec{z}_0$, magnetic
fields when the electron cyclotron frequency, $\omega_c=eB_0/m\gamma$, is
larger than its undulator bounce-frequency, $\omega_u=2\pi v_\parallel/d$.
Here, $e$, $m$, $\gamma$ and $v_\parallel$ are the electron charge, mass,
Lorentz-factor, and longitudinal velocity, and $d$
is the undulator period. In this over-resonance region of parameters, the
increase of the energy, $mc^2\gamma$, of a front electron under the action of
Coulomb field of the rest of the bunch causes approaching of the electron to
the resonance, $\omega_c\approx\omega_u$. If the electrons move along their
quasi-stationary helical trajectories, it is accompanied with increase of the
transverse electron velocity, $v_\perp=cK/\gamma\Delta$, where $K=eB_ud/2\pi
mc$ and $\Delta=1-\omega_c/\omega_u$ are the undulator parameter and the
resonance mismatch.  In certain conditions, the transverse velocity grows so
quickly that the derivative $\partial v_\parallel/\partial\gamma$ is negative,
and the longitudinal components of the front bunch particles velocities
decrease. Simultaneously, particles from the rear part of the bunch lose their
energy and move away from the resonance, which is accompanied with the decrease
of transverse and increase of longitudinal components of velocities. Thus,
similarly to the classic cyclotron Negative-Mass Instability in cyclic
accelerators
\cite{Nielsen_RevSciInst_1959_Longitudinal, Kolomensky_AtomEner_1959_Stability}
and gyrodevices
\cite{Bratman_TechPhys_1976_Instability, Bratman_PhysPlas_1995_‘‘Phase,
Savilov_PhysPlasm_1997_Negative},
the change of the electrons
energy in the repulsing Coulomb field leads to an effective longitudinal
attraction and bunching of non-isochronously oscillating particles.

For typical parameters of a THz source, considered in
\cite{Balal_ApplPhysLett_2015_Negative, Lurie_PhysRevST-A_2016_Energy}
(particle energy of 5--6~MeV, bunch charge of up to 1~nC, and radiation
frequency of 1--2~THz), the value of undulator field, $B_u$, should be of the
order of 0.1--0.2~T, that is tens times smaller than the guiding field, $B_0$.
Correspondingly, a relatively thin steel helix is supposed to be sufficient for
providing the required value of undulator field.

\begin{figure}
     \includegraphics[width=3in]{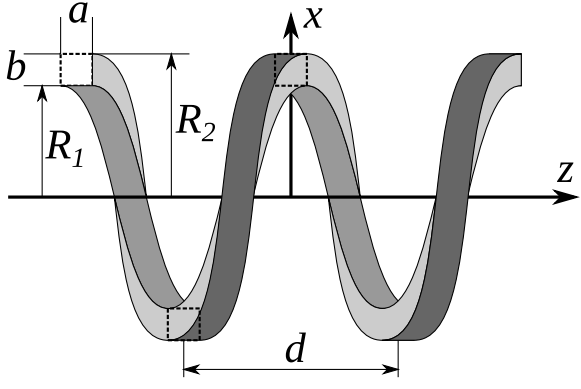}%
 \caption{Ferromagnetic helical insertion into solenoid (two periods are shown).
     \label{schematic}}
 \end{figure}

Let us find the field of a helical insertion placed into an infinite solenoid
with a uniform field. Consider magnetization $\vec{J}$ of an infinite
ferromagnetic helix with period (step) $d$, inner and outer radii $R_1$ and
$R_2$, respectively, thickness $b=R_2-R_1$, and axial size $a$ (Fig.~\ref{schematic}).  In the
case of a very strong guiding field, the magnetization of the helix is saturated,
$\vec{J}=\vec{J}_\infty$, so that its direction practically coincides with direction of
the field, $\vec{z}_0$, and its value does not depend on $B_0$. 

Let us consider an elementary thin helix, for which $R_1\approx R_2\approx R$,
$b\ll R$, and introduce its surface magnetization. Taking into account that the
helix bulk is periodic in axial direction and invariant to the transformations
keeping $\varphi+2\pi z/d=const$, we use the Fourier expansion and
represent the magnetization in the form \begin{equation}
    \vec{J}(r,\varphi,z)=bJ_\infty\delta(r-R)\vec{z}_0\sum_{n=-\infty}^{\infty}
f_ne^{in(hz+\varphi)}, \end{equation} where $h=2\pi/d$ and $f_n=(1/\pi
n)\sin(nha/2)$. The $n$th harmonic of the magnetization, $\vec{J}_n$, presents
a continuous polarized cylindrical surface with radius $R$;
$\mathrm{div}\vec{J}_n$ can be interpreted as equivalent continuously
distributed surface magnetic charge. The corresponding harmonic of the static
magnetic potential satisfies the Poisson equation 
\begin{equation}
    \Delta\psi_n=\mathrm{div}\vec{J}_n.
    \label{eq:lapl}
\end{equation}
Boundary conditions at
the cylindrical surface $r=R$ correspond to continuity of the potential and jump of
the radial component of magnetic field strength $\vec{H}_n=-\nabla\psi_n$:
\begin{equation}
    \left[\psi_n\right]_{r=R}=0,
    \quad\left[H_r\right]_{r=R}=-inhbRJ_\infty f_ne^{in(hz+\varphi)}.
    \label{eq:bcond}
\end{equation}
The solution of Eqs. (\ref{eq:lapl}), (\ref{eq:bcond}) is
\begin{equation}
    \psi_n=-ing_n e^{in(hz+\varphi)}\left\lbrace
    \begin{array}{ll}
        \mathrm{K}_p(phR)\mathrm{I}_p(phr), & r\leq R \\
        \mathrm{I}_p(phR)\mathrm{K}_p(phr), & r\geq R \\
    \end{array}
    \right.,
    \label{eq:sol}
\end{equation}
where $\mathrm{I}_p$, $\mathrm{K}_p$ are modified Bessel and McDonald
functions of the order of $p=|n|$ and $g_n=hbRJ_\infty f_n$.

It is easy to show that only harmonics $n=\pm1$ in Eq.  (\ref{eq:sol})
contribute to the transverse undulator field at the axis of the system. The
corresponding field is
\begin{equation}
    \vec{B}_u(r=0)=\Re\left[\left(\vec{x}_0+i\vec{y}_0\right)B_ue^{i(hz-\pi/2)}\right],
    \label{eq:fvec}
\end{equation}
where
\begin{equation}
    B_u=\frac{1}{\pi}h^2Rb\sin\left(ha/2\right)\mathrm{K}_1(hR)\mu_0J_\infty.
    \label{eq:fsc}
\end{equation}
Maximum of the transverse field is
achieved when the longitudinal size of the helix is equal to half of its
period, $a=d/2$. Integration of expression (\ref{eq:sol}) over the radius gives solution for the helix
of finite thickness.

\section{Numerical simulations}

\begin{figure}
 \includegraphics[width=3in]{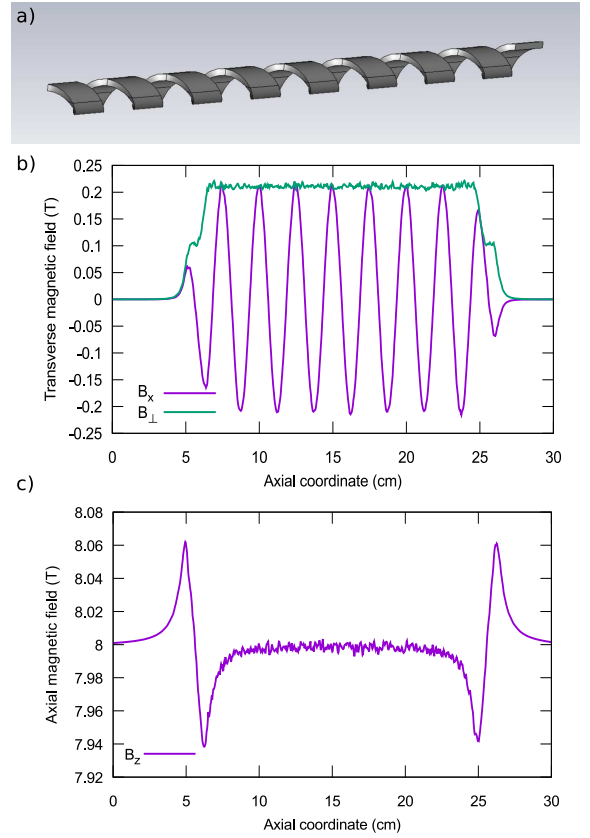}%
    \caption{Results of simulations within the {\it CST Studio} for the case
of steel helix with rectangular cross-section having the parameters of $d=2.5$~cm,
    $a=1.2$~cm, $R_1=0.3$~cm
and $b=0.2$~cm. The helix geometry (a) and the amplitudes of the transverse
(b) and axial (c) magnetic fields.
 \label{cst}}
\end{figure}

The applicability of the solution (\ref{eq:fvec})--(\ref{eq:fsc}) was verified
by direct solving of magnetostatic equations using {\it CST Studio}. This code
allows one to take into account a finite thickness, arbitrary transverse and
longitudinal profiles of the helix as well as nonlinear dependence of magnetic
permeability of steel on applied field. Numerical simulations demonstrate a
high accuracy of the solution (\ref{eq:fvec})--(\ref{eq:fsc}) for long
enough helices. For instance, the estimation (\ref{eq:fsc}) yields for the case
of an iron helix having the parameters of $d=2.5$~cm, $a=1.2$~cm, $R_1=0.3$~cm
and $b=0.2$~cm the value of the transverse magnetic field of 0.24~T, which
admits quite well with 0.21~T given by the {\it CST Studio} simulations for
these parameters and axial magnetic field of $8\mbox{ T }\gg
\mu_0J_\infty\approx2.15$~T (Fig.~\ref{cst}).

\begin{figure}
 \includegraphics[width=3in]{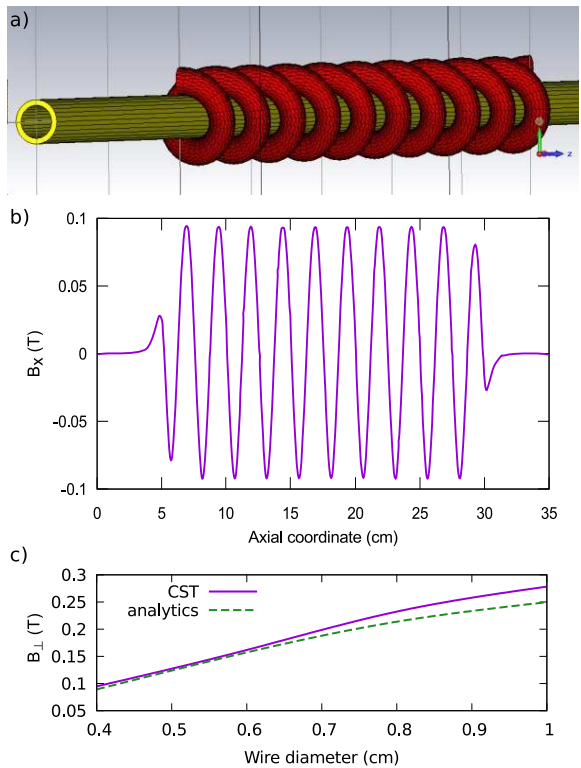}%
    \caption{Results of simulations within the {\it CST Studio} for the case
of iron helix with round cross-section, period of 2.5 cm, and inner radius of 5 mm.
    The helix geometry (a) and the amplitude
    of the transverse magnetic field as function of axial coordinate (b, wire diameter is
    0.4~cm) and of helix wire diameter (c, maximum field amplitude). The predictions
    of formula (\ref{eq:fsc}) for the equivalent square cross-section are also shown (c).
 \label{cst2}}
\end{figure}

Also, formula (\ref{eq:fsc}) can be well applied for thin helices with round cross-sections
if one takes parameters $a$ and $b$ equal to each other, so that $ab$ corresponded to
the square of the wire cross-section, and $R$ equal to the mean radius of the helix bulk
(Fig.~\ref{cst2}).

\section{Experimental measurements}

\begin{figure}
 \includegraphics[width=3in]{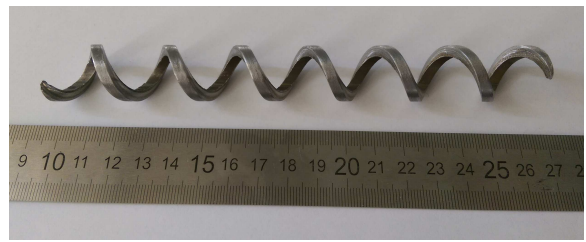}%
    \caption{Steel helix with square cross-section used for experimental
    measurements of induced transverse magnetic field.
 \label{hlx}}
\end{figure}

The described method of creation of helical undulator field has been also
studied experimentally for a helix of square cross-section
(Fig.~\ref{hlx}). In the experiment, using a probe in the form of a small
coil, the transverse field was measured on the axis of the helix placed into
pulsed fields with maximum values of 4--5~T. The probe coil contained $\sim$200
turns and had a length of 5~mm and an average diameter of 5~mm. To compensate
for the inevitable deviation of the coil axis from the radial direction,
leading to a contribution from the strong longitudinal field, the probe was
rotated around the main solenoid ($z$) axis, and a constant component was
subtracted from the measured value.

\begin{figure}
 \includegraphics[width=3in]{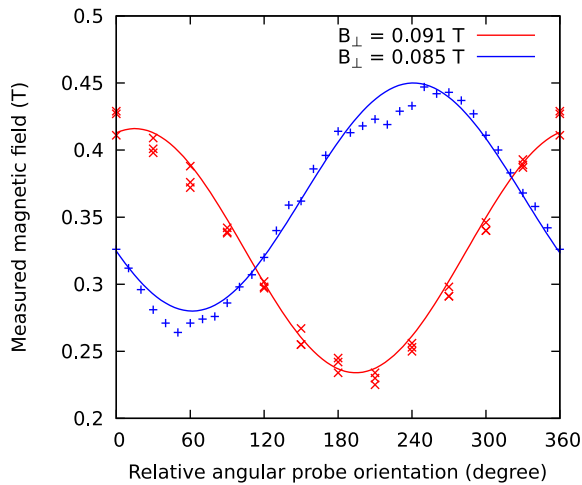}%
\caption{Results of experimental measuring of a transverse magnetic field at the
axis of the steel helix of a square cross-section. The different curves
correspond to different axial positions of the probe. The transverse component
of the field is determined as an amplitude of the sin-like dependency of the
probe signal on its angular position.
 \label{exp}}
\end{figure}

In the first experiment, we used a steel helix with the period of 2.5~cm, inner
radius of 0.6~cm and square cross-section having thickness of 0.4~cm
(Fig.~\ref{hlx}a). The steel type was equivalent to ASTM-SAE 1045 having the
carbon content less than 0.5\%, which means that the saturation magnetic field
was about 2~T \cite{Burrows_Bulletin_1916_Corellation}. The helix was
manufactured from steel cylinder first by milling the cylinder surface and then
by cutting away the inner part by electrical discharge machining. Being mounted
in the pulsed solenoid with an axial magnetic field of about 4.5~T, it created
the transverse helical field of the order of 0.1~T (Fig.~\ref{exp}). The main
source of measurements error was connected with displacements of the probe coil
from the solenoid axis, which were caused partially by misalignment of the
probe support during the rotation of the probe, and partially by mechanical
shakes caused by pulsed magnetic field of large amplitude. The difference in
field values measured at different axial probe positions can be explained by
the axial inhomogeneity of the solenoid field, which had the axial scale of
about 10~cm. It should be noted, that in real application, such a helix is
supposed to be placed inside a DC or long-pulse solenoid, whereas in the described
experiment, the solenoid pulse had an almost sinusoidal form, $B_0\sim\sin(\pi
t/T)$, with a length of ''half-period`` $T\approx3$~ms. In principle, in the
pulsed case, a second source of transverse magnetic field appears, which is the
induction current along the helix.  Nevertheless, simple estimations show that
for these particular pulse duration, helix geometry and material conductance,
the correspondent magnetic field is almost two orders of magnitude lower than the
one induced by magnetization. 

It is very important, that the simple formulas (\ref{eq:fvec})--(\ref{eq:fsc})
give for the experimental parameters the very same value of transverse magnetic
field, $B_u\approx0.09$~T, and so do the {\it CST} numerical simulations. This
proves the correctness of theoretical calculations and assures that the
undulator field of 0.2~T, which was predicted by calculations of Section III
and corresponds to desirable undulator parameter, $K\approx0.5$, indeed can be
achieved experimentally.

\section{Conclusion}

The proposed method of creating a helical undulator field by redistributing the
uniform field at a steel helix seems very simple and efficient. It can be
naturally applied to THz sources of coherent spontaneous undulator radiation
using the NMI stabilization of longitudinal sizes of short picosecond and
sub-picosecond electron bunches \cite{Balal_ApplPhysLett_2015_Negative,
Lurie_PhysRevST-A_2016_Energy}. According to simulations, development of NMI in
a combined strong uniform and helical undulator fields can provide maintenance
of nearly constant sizes of short and long-living bunch cores at a fairly large
undulator length. It makes possible the implementation of powerful and
narrowband radiation sources with initial electron energy of about 5--6~MeV and
charge of the order of 1~nC with the frequency of 1--2~THz and efficiency of
about of 20\%. 

\begin{acknowledgments}
The authors are grateful to A.~V.~Savilov for useful discussions. The work
is supported by Russian Foundation for Basic Research, project 16-02-00794, and
by Israeli Ministry of Science, Technology and Space.
\end{acknowledgments}

%\bibliography{bibfile}
%

\end{document}